\documentclass[conference]{IEEEtran}
\IEEEoverridecommandlockouts

\usepackage{xcolor}
\usepackage{multirow}
\usepackage{algorithm}
\usepackage{algpseudocode}
\usepackage{amsmath}
\usepackage{algorithmicx}
\usepackage{graphicx}
\usepackage{booktabs}
\usepackage{hyperref}
\usepackage{cite}
\newcommand{\lbname}{Spritz }
\newcommand{\lbnamed}{Spritz}

\def\BibTeX{{\rm B\kern-.05em{\sc i\kern-.025em b}\kern-.08em
    T\kern-.1667em\lower.7ex\hbox{E}\kern-.125emX}}

\begin{document}

\title{Spritz: Path-Aware Load Balancing in Low-Diameter Networks}

\author{
\IEEEauthorblockN{
Tommaso Bonato\IEEEauthorrefmark{1}\IEEEauthorrefmark{2}\IEEEauthorrefmark{3},
Ales Kubicek\IEEEauthorrefmark{1}\IEEEauthorrefmark{3},
Abdul Kabbani\IEEEauthorrefmark{2},
Ahmad Ghalayini\IEEEauthorrefmark{2},
Maciej Besta\IEEEauthorrefmark{1},
Torsten Hoefler\IEEEauthorrefmark{1}\IEEEauthorrefmark{2}
}
\IEEEauthorblockA{
\IEEEauthorrefmark{1}ETH Z\"urich, Switzerland \quad
\IEEEauthorrefmark{2}Microsoft, USA
}
\IEEEauthorblockA{
Emails:
\texttt{\{tommaso.bonato,maciej.besta,torsten.hoefler\}@inf.ethz.ch},
\texttt{akubicek@student.ethz.ch},\\
\texttt{\{abdulkabbani,aghalayini\}@microsoft.com}
}
\IEEEauthorblockA{
\IEEEauthorrefmark{3}\emph{Both authors contributed equally to this research.}
}
}

\maketitle

\begin{abstract}
Low-diameter topologies such as Dragonfly and Slim Fly are increasingly adopted in HPC and datacenter networks, yet existing load balancing techniques either rely on proprietary in-network mechanisms or fail to utilize the full path diversity of these topologies. We introduce Spritz, a flexible sender-based load balancing framework that shifts adaptive topology-aware routing to the endpoints using only standard Ethernet features. We propose two algorithms, Spritz-Scout and Spritz-Spray that, respectively, explore and adaptively cache efficient paths using ECN, packet trimming, and timeout feedback. Through simulation on Dragonfly and Slim Fly topologies with over 1000 endpoints, Spritz outperforms ECMP, UGAL-L, and prior sender-based approaches by up to 1.8x in flow completion time under AI training and datacenter workloads, while offering robust failover with performance improvements of up to 25.4x under link failures, all without additional hardware support. Spritz enables datacenter-scale, commodity Ethernet networks to efficiently leverage low-diameter topologies, offering unified routing and load balancing for the Ultra Ethernet era.
\end{abstract}

\section{Introduction}\label{sec:introduction}

High-performance computing (HPC) clusters and modern datacenters have traditionally operated in distinct domains, driven by different workloads, performance targets, and networking philosophies. HPC clusters prioritize ultra-low latency and high throughput for tightly-coupled scientific applications, often employing custom high-performance interconnects~\cite{yin2022comparative, hoefler2022convergence}. In contrast, datacenters are designed to serve a wide array of services with reliability and elasticity in mind, relying heavily on commodity Ethernet and standardized protocols~\cite{yin2022comparative, hoefler2022convergence}.

However, in recent years, HPC clusters and datacenters have begun to converge~\cite{hoefler2022convergence}. As modern datacenters increasingly handle HPC-style workloads such as large-scale AI training and inference, the performance and scalability demands of both environments are becoming aligned. This shift is one of the main motivations behind efforts like Ultra Ethernet (UE)~\cite{ultraethernet2025spec}, which aims to bring HPC-grade performance into Ethernet-based systems and enable a unified, high-performance, and scalable network architecture~\cite{ue}.

As Ethernet-based systems increasingly support large-scale distributed workloads such as AI training, the importance of effective load balancing in multi-tenant environments continues to grow. Traditional routing schemes like Equal Cost Multi-Path (ECMP) routing~\cite{hopps2000analysis} struggle under uneven traffic distributions, link failures, and from hashing collisions~\cite{zhang2021hashing}. This often leads to congestion, packet drops, and retransmissions~\cite{hoefler2023data}. Studies have shown that even a single link failure can significantly degrade training performance and increase costs~\cite{gangidi2024rdma, qian2024alibaba}. These issues highlight the need for load balancing strategies that respond quickly to shifting network conditions while preserving compatibility with existing Ethernet infrastructure.

\begin{figure}[!t]
\centering
\includegraphics[width=1.0\linewidth]{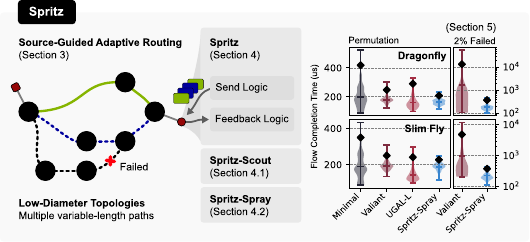}
\caption{
Overview of the Spritz framework.
}
\label{fig:poster}
\vspace{-1.5em}
\end{figure}

Much of the existing research on congestion control and load balancing~\cite{katta2016hula, vanini2017let, alizadeh2014conga, katta2016clove, he2015presto, zhang2017resilient, huang2020tuning, alasmar2018polyraptor, montazeri2018homa, hwang2014deadline, lu2016sed, vamanan2012deadline, alizadeh2013pfabric, bai2014pias, raiciu2011improving, handley2017re, zhuo2016rack, he2016ac, alizadeh2010data, cardwell2017bbr, mittal2015timely, li2013openflow, perry2014fastpass, olteanu2016datacenter, park2019maxpass, benet2018mp, lu2018multi, widmer2001survey, callegari2014survey, bonato2025flowcutswitchinghighperformanceadaptive} in datacenter networks, including the recently released Ultra Ethernet specifications~\cite{hoefler2025ultraethernetsdesignprinciples}, focuses on fat tree topologies. These networks offer many equal-length paths between endpoints, which aligns with ECMP routing. This key property of fat tree topologies makes it easier to design sensible load balancing solutions as there are only minimal paths between any two endpoints.


However, as of August 2025, seven of the ten most powerful HPC clusters use a Dragonfly-based topology (the HPE Slingshot-11 interconnect)~\cite{top5002024}. Moreover, systems using HPE Slingshot-11 account for 53\% of the total performance share in the Top500 list with an increasing trend in recent years. Unfortunately, although Slingshot-11 is Ethernet-based, it relies on proprietary adaptive routing and load balancing features implemented in custom Rosetta switches~\cite{de2020depth, rowethhpe}, limiting its accessibility and general applicability. On top of that, many other high-performance commodity Ethernet switches offer some form of adaptive routing for in-network load balancing. However, there is no unified standard and therefore these features are vendor specific and with few public details (Broadcom DLB/GLB and Cognitive Routing, Juniper DLB/GLB, NVIDIA Adaptive RoCE)~\cite{broadcom2025routing, juniper2025routing, nvidia2025routing, brodong2025draft}. This limits interoperability, complicates deployment, and restricts vendor diversity within or across datacenters. The issue is further aggravated by the complexity of routing in low-diameter networks such as Dragonfly or Slim Fly, where achieving high performance requires the use of \textit{both minimal and non-minimal} paths~\cite{besta2020fatpaths}.

To address this gap, we shift adaptive routing decisions from network switches to source endpoints by introducing Source-Guided Adaptive Routing (\textbf{contribution~\#1}). Our approach requires no major hardware changes and only standard ECMP support, which is available in commodity Ethernet switches and avoids reliance on proprietary vendor-specific features. We focus on Dragonfly, which is widely deployed in production systems~\cite{kim2008technology, top5002024}, and Slim~Fly, which has at least one known deployment and offers strong scalability properties~\cite{besta2014slim, blach2024high}. Next, we introduce a framework for the design of congestion control and load balancing algorithms in low-diameter topologies while maintaining compatibility with existing Ethernet infrastructure (\textbf{contribution~\#2}). Using this framework, we introduce two novel sender-based load balancing algorithms, \lbnamed-Scout and \lbnamed-Spray (\textbf{contribution~\#3}). Finally, we evaluate these algorithms through simulation on both Dragonfly and Slim~Fly topologies to demonstrate their effectiveness (\textbf{contribution~\#4}). Figure~\ref{fig:poster} outlines our contributions.
\section{Background}\label{sec:background}
This section provides background on network topologies,     routing strategies, and sender-based load balancing. We review key design principles of fat tree and low-diameter topologies. We then outline common routing strategies. Finally, we describe sender-based techniques for efficient traffic distribution.

\subsection{Network Topologies}\label{sec:background_topologies}
\paragraph{Fat Tree Topology} 
Fat tree topologies are widely used in modern datacenters. Their regular and symmetric structure makes them easy to deploy and manage at scale. Most configurations use either two or three levels of switches. In this hierarchy, links used to traverse the network upwards are referred to as up-links, while links going towards lower layer switches are referred to as down-links.

A fully provisioned fat tree offers full bisection bandwidth and supports multipathing. However, it is expensive to build at scale~\cite{hoefler2022hammingmesh}. To reduce cost, many deployments oversubscribe up-links to save cost on both cabling and switches.  Several established routing schemes exist for fat tree topologies:

\textbf{Equal Cost Multi-Path (ECMP)} routing selects the next hop by hashing fields in the packet header, typically: source and destination IPs, ports, and protocol number. Based on this hash, a specific up-link path is chosen from the ECMP table, and the packet is forwarded along that path. This scheme is widely supported on commodity Ethernet switches and is usually used because it is simple and guarantees in-order delivery of packets from the same message.

\textbf{Adaptive} routing selects the next hop based on in-network information, such as local queue occupancy. These decisions aim to avoid congestion by steering traffic toward less congested paths. This often relies only on local information without having a global view of the network. Moreover, such schemes often rely on proprietary switch features that are not standardized across vendors~\cite{brodong2025draft}.

\paragraph{Low-Diameter Topologies} 

Low-diameter topologies are common in modern HPC clusters. They are designed to reduce the number of hops between any two switches, which lowers latency and improves bandwidth~\cite{kim2008technology, besta2014slim}. Unlike fat trees, they can scale with fewer switches and cables, making them attractive for large systems with strict cost and performance requirements~\cite{kim2008technology, besta2014slim}.

Several low-diameter topologies have been proposed in recent years~\cite{kim2008technology, ahn2009hyperx, besta2014slim, lakhotia2022polarfly, lakhotia2024polarstar, lei2020bundlefly}. In this work, we focus on two representative designs that have been deployed in practice: Dragonfly (diameter 3)~\cite{kim2008technology} and Slim Fly (diameter 2)~\cite{besta2014slim}.

A key difference from fat trees is that low-diameter topologies offer both minimal and non-minimal paths between endpoints. This introduces a trade-off between path length and congestion. Choosing between a shorter but congested path and a longer but less congested one makes routing and load balancing more complex. Moreover, switches are organized into groups, with short local links connecting switches within a group and longer global links connecting switches across groups~\cite{kim2008technology, besta2014slim}. The latency difference between local and global links is significant and is analyzed in Section~\ref{sec:routing}.  Several routing schemes have been developed to address this:

\textbf{Minimal} routing always selects the next hop only along a minimal path between endpoints. It is simple and efficient under uniform traffic but performs poorly under adversarial patterns due to limited path diversity~\cite{kim2008technology, besta2014slim}. It can be implemented using static routing on Ethernet switches.

\textbf{Valiant} routing addresses the limitations of minimal routing by first selecting a random intermediate location, then routing minimally to the final destination~\cite{valiant1982scheme, kim2008technology, besta2014slim}. This randomization spreads traffic across the network and improves path diversity. However, it introduces additional latency due to non-minimal paths and offers no control over path selection.

\textbf{UGAL-L} routing combines the benefits of minimal and non-minimal routing by making per-packet decisions based on estimated congestion. It compares the cost of a minimal path with that of a valiant (non-minimal) path and selects the one with lower estimated load~\cite{kim2008technology, besta2014slim}. It relies only on local information which can lead to suboptimal global decisions. UGAL-G improves this by incorporating global congestion estimates, but requires additional communication and system complexity~\cite{kim2008technology, besta2014slim}. Both variants require proprietary switch features that are not standardized across vendors.


\subsection{Congestion Signals}\label{sec:background_signals}

\sloppy{
\paragraph{Explicit Congestion Notification (ECN)} ECN allows switches to signal congestion by marking a bit in the IP header. The receiver returns this signal in the ACK, providing feedback to the sender about network congestion. Switches typically mark packets based on queue size using schemes like Random Early Detection (RED), where marking probability increases linearly between defined thresholds ($K_{min}$ and $K_{max}$)~\cite{floyd1993random}. ECN is widely supported in commodity Ethernet switches, and has been deployed at large scale~\cite{singh2015jupiter, zhu2015congestion}.
}

\paragraph{Packet Trimming} Packet trimming allows switches to remove the payload of a packet instead of dropping it entirely when reaching a certain threshold (which could be equal to the buffer size). This preserves the header, enabling the receiver to detect and report severe congestion to the sender. It provides faster congestion feedback and is being adopted in the emerging Ultra Ethernet standard~\cite{adrian2022implementing, ueupdate}.

\paragraph{Timeout} Timeouts are used to detect packet loss and are often treated as a signal of severe congestion or failure. While they react slowly and are difficult to tune, timeouts serve as a last resort when no other feedback is available. In systems with packet trimming, a timeout may indicate that a packet was dropped entirely, such as when a switch silently discards it. This makes timeouts useful for detecting failures like black holes in the network.

\subsection{Sender-Based Load Balancing}\label{sec:background_lb}

In our discussion, we distinguish between several core concepts in the context of sender-based load balancing. 

\paragraph{Message} \textit{Message} is a high-level object sent by the programmer that encapsulates the intended data or command. \textit{Flow} is a sequence of packets that collectively implements the transmission of the message. This distinction allows us to describe communication at different levels of abstraction. \textit{Flowlet} is a burst of packets within a flow, typically separated by short idle periods. Flowlets enable finer-grained load balancing, as they can follow different network paths~\cite{vanini2017let, kandula2007dynamic}.

\paragraph{ECMP Hashing} Modifying one of these fields (introduced in Section~\ref{sec:background_topologies}) can change the hash and redirect traffic. Since ECMP ignores network load, it may map many flows to the same path, leading to collisions~\cite{al2010hedera, alizadeh2014conga}.

\paragraph{Entropy Value (EV)} EV refers to a header field, often the source port, that can be safely modified to influence the ECMP hashing. Changing this value likely alters the selected path without affecting application semantics. Due to hash collision, two same EVs might still map to the same path. 


\section{Source-Guided Adaptive Routing}\label{sec:routing}

\begin{figure*}[!ht]
\centering
\includegraphics[width=1.0\textwidth]{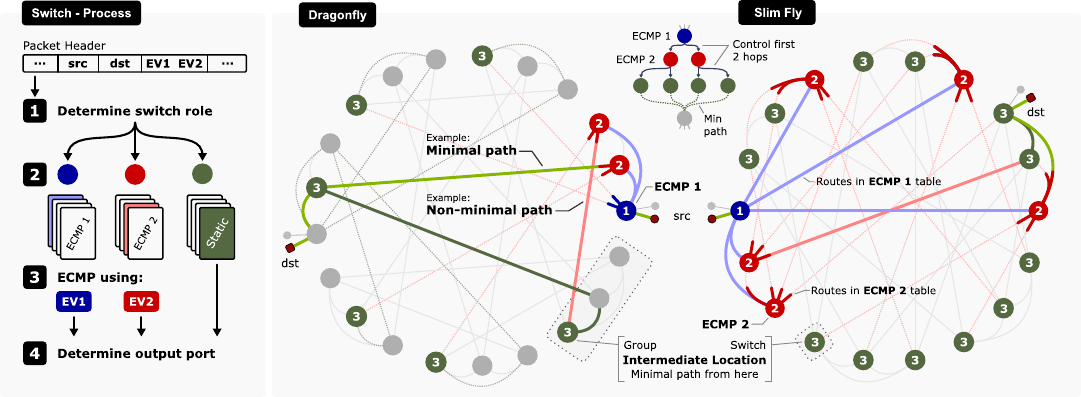}
\caption{
Decision logic on each switch (left). Source-Guided Adaptive Routing in Dragonfly and Slim Fly (right). The routing process from a source endpoint \includegraphics[scale=0.6,trim=0 1 0 0]{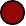} to a destination endpoint \includegraphics[scale=0.6,trim=0 1 0 0]{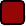} involves at most three steps. \includegraphics[scale=0.6,trim=0 1 0 0]{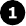}~Control of the first hop at the ECMP 1 switch \includegraphics[scale=0.6,trim=0 1 0 0]{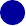} using EV1. \includegraphics[scale=0.6,trim=0 1 0 0]{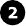}~Control of the second hop at the ECMP 2 switch \includegraphics[scale=0.6,trim=0 1 0 0]{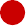} using EV2 (this switch is always a direct neighbour of the ECMP 1 switch). \includegraphics[scale=0.6,trim=0 1 0 0]{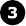}~Reaching an intermediate location \includegraphics[scale=0.6,trim=0 1 0 0]{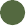}. From there, default forwarding table is used to reach the destination. The source uses EV1 and EV2 to guide the routing decisions. Based on these values, the path can be either minimal (\includegraphics[scale=0.6,trim=0 -2 0 0]{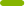}) or non-minimal (\includegraphics[scale=0.6,trim=0 -2 0 0]{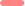} + \includegraphics[scale=0.6,trim=0 -2 0 0]{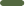}) with various length. 
}
\label{fig:routing}
\vspace{-1.2em}
\end{figure*}

In this section, we present a routing scheme that shifts path selection from switches to endpoints in low-diameter topologies such as Dragonfly and Slim Fly. The key insight is that by controlling up to the first two hops, the source can reach all bounded simple paths between any source and destination. We define bounded simple paths as those with up to 3 local and 2 global hops in Dragonfly, and up to 4 global hops in Slim Fly. These bounds follow from the definition of Valiant routing in both topologies. In other words, we target all possible Valiant paths between any source and destination. Table~\ref{tab:path_categories} shows all possible types of bounded simple paths in our setting.

\paragraph{Dragonfly} Valiant routing selects an intermediate group as the misroute target~\cite{kim2008technology}. Within each group, all switches are connected to all other switches. In addition, every group has at least one global link to every other group. As a result, any intermediate group can be reached in at most two hops from the source. We focus primarily on the original Dragonfly design with all-to-all intra-group connectivity. However, our scheme is extendable to Cray Cascade (2D all-to-all intra-group)~\cite{faanes2012cray} or Dragonfly+ (Clos-like intra-group)~\cite{shpiner2017dragonfly+} variants.

\paragraph{Slim Fly} Valiant routing selects an intermediate switch as the misroute target~\cite{besta2014slim}. Since the topology has a diameter of two, any intermediate switch can be reached in at most two hops from the source.\\

We assume that switches support features commonly available in commodity Ethernet hardware. These include Virtual Routing and Forwarding (VRF), multiple ECMP tables, and configurable ECMP hash functions~\cite{zhang2021hashing}. 

For the rest of this work, we use the following terminology:

\paragraph{Entropy Value (EV)} 16-bit value placed in the packet header (Section~\ref{sec:background_lb}) to influence ECMP routing. We split EV into two parts: EV1 (most significant 8 bits) and EV2 (least significant 8 bits). Switches use either EV1 or EV2 in ECMP hashing, depending on their role for each packet (Section~\ref{sec:routing_ecmp}). As we assume full control of the switch hash function, we will treat each EV as a unique path between a particular source and destination.

\paragraph{EV Entry} Source-side data structure containing EV (EV1, EV2), and associated statistics (size: 16 + 8 bits).

\paragraph{EV Entry List} Source-side list of one or more EV entries. The aim is to enumerate unique paths between that source and a particular destination.

\paragraph{Endpoint Table} Source-side key-value lookup table. Stores an EV entry list per destination switch. We apply static compression and use the destination switch as the key, since all endpoints behind the same switch share the same list. Organization of this table, along with memory requirements is discussed in Section~\ref{sec:routing_endpoints}.

\subsection{Switch Role and ECMP Hashing}\label{sec:routing_ecmp}

We assume that each switch has multiple pre-configured ECMP tables and a single default forwarding table with static minimal routes. The switch decision logic is shown in Figure~\ref{fig:routing}. 

\includegraphics[scale=0.7,trim=0 1 0 0]{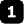} \textbf{Role Detection} The switch is programmed to determine its role based on its position along the packet path. It identifies as: \textit{ECMP~1},  \textit{ECMP~2}, or an \textit{intermediate location}. 

\includegraphics[scale=0.7,trim=0 1 0 0]{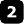} \textbf{Table Choice} Based on the detected role, the switch selects either an ECMP table (roles: ECMP~1 or ECMP~2) or the default forwarding table (role: intermediate location).

\includegraphics[scale=0.7,trim=0 1 0 0]{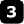} \textbf{ECMP} If the switch is in the ECMP~1 or ECMP~2 role, it performs ECMP routing. EV1 is used as input to the hash function for ECMP~1, and EV2 for ECMP~2. In the fine-grained version of our scheme, we assume the switch uses an identity hash function with modulo table size, allowing endpoints to precisely track individual paths. The coarse-grained version, without this requirement, is described in Section~\ref{sec:routing_coarse}.

\includegraphics[scale=0.7,trim=0 1 0 0]{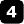} \textbf{Route} Based on the selected entry from the ECMP table or the default forwarding table, the packet is forwarded to the corresponding output port.

\subsection{Routing Process}

We now describe the routing process from the perspective of a packet traversing the network. It consists of three initial steps that enable routing along any bounded simple path. This process is illustrated in Figure~\ref{fig:routing}.

\includegraphics[scale=0.6,trim=0 1 0 0]{figures/icons/Circle1Icon.pdf}
\textbf{ECMP 1} The switch identifies its role as ECMP 1 based on the packet header. It activates the corresponding ECMP table and uses entropy value \textbf{EV1} in the hash function to select the first hop. In Dragonfly, if the source and destination are in the same group, a different ECMP table is used. This avoids misrouting outside the group, which would violate the bounded simple path property.

\includegraphics[scale=0.6,trim=0 1 0 0]{figures/icons/Circle2Icon.pdf}
\textbf{ECMP 2} Similar to ECMP 1, but the switch uses entropy value EV2 to select the second hop. In Dragonfly, if the source and destination are in the same group, it uses the default forwarding table instead of an ECMP table.

\includegraphics[scale=0.6,trim=0 1 0 0]{figures/icons/Circle3Icon.pdf}
\textbf{Intermediate Location} The packet reaches an intermediate location. This is a single switch in a remote group in Dragonfly or an individual switch in Slim Fly. Then switches use default forwarding to route on a \textbf{minimal path}.

Switches without a role forward via the default table on a minimal path.

Spritz does not alter the underlying deadlock-free routing/VC (Virtual Channel) scheme; EV1/EV2 only select among VC-safe next hops. In Dragonfly, the switch uses EV1 and EV2 to reach at most one intermediate group and then reverts to minimal routing; thus any non-minimal path crosses a single intermediate group.

\subsection{Endpoints}\label{sec:routing_endpoints}

\begin{figure}[!t]
\centering
\includegraphics[width=1.0\linewidth]{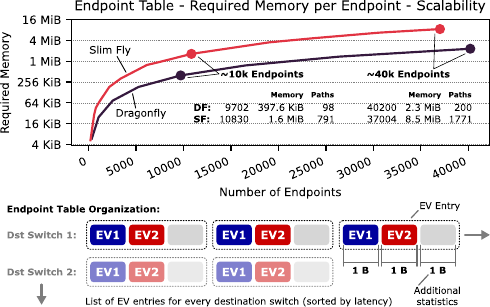}
\caption{
Upper bound on the memory required per endpoint to store the endpoint table. For each destination switch, the source endpoint stores an EV entry list, enumerating bounded simple paths to reach that destination (sorted by latency). We show how the requirements scale for Dragonfly and Slim Fly topologies up to 40k endpoints when enumerating all bounded simple paths. \textit{Paths} refers to the maximum number of paths for any source-destination pair.
}
\label{fig:endpoint_mem}
\vspace{-1.5em}
\end{figure}

Each source endpoint maintains path information for every destination in the network. This information is encoded as EV entries. Each EV entry includes the entropy value (EV1, EV2), which control the hash functions at the ECMP 1 and ECMP 2 switches. In other words, they determine the first two hops of the packet. Each EV1 and EV2 are represented using 8-bits each, allowing for 256 unique values. This allows the hash function during ECMP to select any output port on modern high-radix switches (256 ports). An additional byte is reserved for metadata about the path, such as latency, disjointness, or statistics. For example, if a timeout is observed for a given EV entry, it may be marked as unavailable and avoided in future.

The list of EV entries can be sorted by latency and partitioned by path categories or further by path types. As shown in Table~\ref{tab:path_categories}, there are significant latency differences not only between minimal and non-minimal paths, but also among non-minimal paths themselves, especially in Slim Fly.

The main limitation of encoding all possible paths on the endpoint is the additional memory required to store the endpoint lookup table. Figure~\ref{fig:endpoint_mem} shows the memory required for both Dragonfly and Slim Fly topologies with up to 40k endpoints. At this scale, the required memory is still reasonable at approximately 2.3 MiB (maximum of 200 bounded simple paths per destination) for Dragonfly and 8.5 MiB (maximum of 1771 bounded simple paths per destination) for Slim Fly.

\begin{table}[t]
\centering
\caption{
Path categories and corresponding path types that may occur between source and destination endpoints. Each path type is defined by the number of local and global hops.
}
\begingroup
\setlength{\tabcolsep}{4pt}      
\renewcommand{\arraystretch}{0.9}
\scriptsize   
\begin{tabular}{@{}l|ccr|ccr@{}}
\toprule
\multirow{2}{*}{\textbf{Path Category}} & \multicolumn{3}{c|}{\textbf{Dragonfly}} & \multicolumn{3}{c}{\textbf{Slim Fly}} \\
\cmidrule(lr){2-4} \cmidrule(lr){5-7}
& \textbf{Local} & \textbf{Global} & \textbf{Latency} & \textbf{Local} & \textbf{Global} & \textbf{Latency} \\
\midrule
Minimal & 1 & 0 & 108.2 ns & 1 & 0 & 108.2 ns \\
(within group) & 2 & 0 & 216.4 ns & 2 & 0 & 216.4 ns \\
\midrule
Minimal & 0 & 1 & 583.2 ns & 0 & 1 & 583.2 ns \\
(across groups) & 1 & 1 & 691.4 ns & 1 & 1 & 691.4 ns \\
 & 2 & 1 & 799.6 ns & 0 & 2 & 1166.4 ns \\
 \midrule
Non-minimal & 0 & 2 & 1166.4 ns & 3 & 0 & 324.6 ns \\
 & 1 & 2 & 1274.6 ns & 4 & 0 & 432.8 ns \\
 & 2 & 2 & 1382.8 ns & 2 & 1 & 799.6 ns \\
 & 3 & 2 & 1491.0 ns & 3 & 1 & 907.8 ns \\
 &  &  &  & 1 & 2 & 1274.6 ns \\
 &  &  &  & 2 & 2 & 1382.8 ns \\
 &  &  &  & 0 & 3 & 1749.6 ns \\
 &  &  &  & 1 & 3 & 1857.8 ns \\
 &  &  &  & 0 & 4 & 2332.8 ns \\
\bottomrule
\end{tabular}
\endgroup
\label{tab:path_categories}
\vspace{-1em}
\end{table}

In deployments where storing the complete endpoint lookup table is not feasible, a subset of EV entries may be selected. Statistical methods to select such subsets can be adapted from FatPaths routing~\cite{besta2020fatpaths}. This presents a tradeoff between memory usage and utilization of all bounded simple paths.



\subsection{Coarse-Grained Path Control}\label{sec:routing_coarse}

In scenarios where fine-grained control of selecting a specific path is not necessary, the source endpoint may mark outgoing packets with a VLAN tag that represents a path category (e.g., minimal or non-minimal). In this case, the switch first activates ECMP table based on the VLAN tag and then selects a specific path within that table using standard ECMP hashing~\cite{kabbani2017flier}. However, since the source does not keep track of which EV corresponds to which path, any path from the corresponding set can be taken. This presents a tradeoff between fine-grained control with more memory and management overhead and coarse-grained control with less memory overhead.

When VLAN tags are used to control the ECMP table selection, a reserved VLAN tag must indicate use of the default forwarding table (empty VLAN tag). To enforce minimal routing from the intermediate location, switches operating in ECMP~1 or ECMP~2 roles must clear the VLAN tag in the packet header before forwarding.
\section{\lbnamed: Load Balancing}\label{sec:loadbalancing}

\begin{figure}[!h]
\centering
\includegraphics[width=1.0\linewidth]{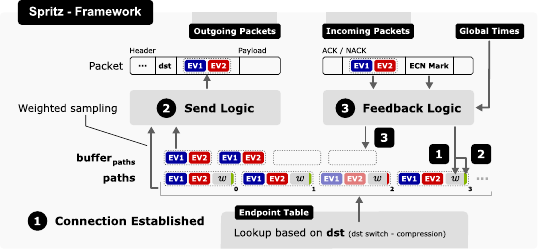}
\caption{
Main components of the \lbname framework. \includegraphics[scale=0.6,trim=0 1 0 0]{figures/icons/Circle1Icon.pdf}~Load a particular EV entry list from the endpoint table when a particular connection is established. \includegraphics[scale=0.6,trim=0 1 0 0]{figures/icons/Circle2Icon.pdf}~Send logic to control the path preference in the network. \includegraphics[scale=0.6,trim=0 1 0 0]{figures/icons/Circle3Icon.pdf}~Feedback logic to gather information about good paths and bad paths.
}
\label{fig:framework}
\vspace{-0.5em}
\end{figure}

Building on the Source-Guided Adaptive Routing described in Section~\ref{sec:routing}, we now introduce a general framework for designing sender-based load balancing and congestion control algorithms for low-diameter topologies. Figure~\ref{fig:framework} illustrates the key components of the control loop. While this work focuses on load balancing, we describe in Section~\ref{sec:loadbalancing_inter} how our approach can be combined with existing sender-based congestion control mechanisms.

We formalize an EV entry as a tuple ${EV_{entry}}_i = ({EV_1}_i, {EV_2}_i, w_i)$, where ${EV_1}_i$ and ${EV_2}_i$ are entropy values used to guide the first two hops, and $w_i$ is a weight used for non-uniform weighted sampling in the send logic. These weights are initialized based on the latency of each path relative to the longest path: 

\begin{equation}\label{eqn:weight}
    w_i = \frac{lat_{longest}}{lat_i}
\end{equation}

For example, if $lat_{shortest} = 799.6$ ns and $lat_{longest} = 1491.0$ ns (based on Table~\ref{tab:path_categories}), then $w_{shortest} = \frac{1491.0}{799.6} \approx 1.86$ and $w_{longest} = 1.0$. This means that the shortest path is almost twice as likely to be selected. The weights can be further scaled using a parameter $w_{scale}$ to increase the preference for shorter paths.

\includegraphics[scale=0.6,trim=0 1 0 0]{figures/icons/Circle1Icon.pdf}~\textbf{Connection Established}
For each connection that is established, the receiver information is part of the message metadata and is used to determine the destination switch ID. Using this ID, the sender retrieves the corresponding EV entry list, referred to as $paths$. Each entry in $paths$ is a 16-bit EV, composed of a tuple $(EV_1, EV_2)$. The sender also initializes a fixed size $\mathit{buffer_{paths}}$ with 8 positions to cache good paths that could be reused. This buffer can be cached to maintain good paths between identical connections with optional reset after some defined time to avoid using stale network information.  

\includegraphics[scale=0.6,trim=0 1 0 0]{figures/icons/Circle2Icon.pdf}~\textbf{Send Logic}
Given a packet $pkt$, the weight vector $w$, $paths$, and $\mathit{buffer_{paths}}$, the sender chooses a path in one of two ways. If the buffer contains entries, a cached good path may be reused. Otherwise, a path index $i$ is sampled using a weighted random generator $gen(w)$ based on the current state of the weight vector. The selected EV is then inserted into the packet header before transmission.

\includegraphics[scale=0.6,trim=0 1 0 0]{figures/icons/Circle3Icon.pdf}~\textbf{Feedback Logic}
Feedback is obtained either by receiving a response packet or by observing the absence of a response within a timeout window. Each sent packet may result in one of the following outcomes:

\begin{itemize} 
    \item \textbf{ACK (no ECN mark):} Good path.
    \item \textbf{ACK (ECN mark):} Congestion is building up.
    \item \textbf{NACK:} Packet was dropped due to full queues.
    \item \textbf{Timeout:} No response was received within a set time interval. Suggests a possible failure along the path. 
\end{itemize}

We assume the receiver keeps the same EV for the response, therefore the sender knows the initial path used for the packet in send logic. 

\begin{itemize} 
    \item \includegraphics[scale=0.7,trim=0 1 0 0]{figures/icons/Square1Icon.pdf}~\textbf{Update weight:} Increase or decrease $w_i$.
    \item \includegraphics[scale=0.7,trim=0 1 0 0]{figures/icons/Square2Icon.pdf}~\textbf{Disable path:} Set $w_i$ to zero.
    \item \includegraphics[scale=0.7,trim=0 1 0 0]{figures/icons/Square3Icon.pdf}~\textbf{Cache path:} Add good path to $\mathit{buffer_{paths}}$. 
\end{itemize}

Using this general framework, we propose two concrete load balancing algorithms for low-diameter topologies: \lbnamed-Scout (Section~\ref{sec:algo_scout}) and \lbnamed-Spray (Section~\ref{sec:algo_spray}). 



\subsection{\lbnamed-Scout}\label{sec:algo_scout}


Algorithm~\ref{algo:send_both} shows the send logic used by both \lbnamed-Scout and \lbnamed-Spray. The two algorithms differ in how they handle the reuse of good paths.

\lbnamed-Scout explores available paths using weighted sampling. Once a path is identified as good, it is reused for subsequent packets until it is removed based on negative feedback. \lbnamed-Spray instead removes each EV from the buffer after it is used. An EV can reenter the buffer only through another positive feedback. 

For both algorithms, we define a configurable parameter \textit{explore\_threshold}. It forces sampling of a new path after every $n$ packets. This ensures periodic exploration during long messages, allowing the sender to discover shorter paths that may have become less congested over time.

\begin{algorithm}[!t]
  \caption{\lbnamed-Scout and \lbnamed-Spray - Send Logic}
  \label{algo:send_both}
  \begingroup
  \footnotesize  
  \begin{flushleft}
    \textbf{Global} weight vector $w$, EV entry list $paths$, cache of good paths $\mathit{buffer_{paths}}$, $packet\_count$, $explore\_threshold$\\
    \textbf{Input:} packet to send $pkt$\\    
  \end{flushleft}
  
  \begin{algorithmic}[1]
    \State \textbf{
      ------------------------------------ Send Logic  ------------------------------------}
    \If{$packet\_count > explore\_threshold$}
      \State $packet\_count \gets 0$
      \State Sample index $i \gets gen(w)$
      \State \Return $paths[i]$
    \EndIf
    \State $packet\_count \gets packet\_count + 1$
    \If{$\mathit{buffer_{paths}}$ is empty}
      \State Sample index $i \gets gen(w)$
      \State \Return $paths[i]$
    \Else
      \State $EV \gets$ front element of $\mathit{buffer_{paths}}$
      \State \textbf{
      ---------------------------- Only \lbnamed-Spray  -----------------------------}
      \State Remove front element from $\mathit{buffer_{paths}}$ 
      \State \textbf{
      -----------------------------------------------------------------------------------}
      \State \Return $EV$
    \EndIf
  \end{algorithmic}
  \endgroup
\end{algorithm}

Algorithm~\ref{algo:feedback_scout} shows the feedback logic used in \lbnamed-Scout. It inserts good paths into $\mathit{buffer_{paths}}$ based on latency, giving priority to shorter paths. Paths are added only if they are not already present in the buffer. The logic also removes paths from the buffer in response to ECN marks, NACKs, or timeouts. In addition, when a timeout occurs, the corresponding weight $w_i$ is temporarily set to zero to prevent further sampling. This value is automatically reset back to the original value (Equation~\ref{eqn:weight}) after a configurable interval (global timer), which can be tuned based on typical failure durations in production. Finally, when we detect uniformly high congestion (i.e., a high ECN-mark rate indicative of large incasts or all-to-all phases), Spritz-Scout biases toward minimal paths, since extra hops only increase delay when all routes are congested.

\begin{algorithm}[!t]
  \caption{\lbnamed-Scout - Feedback Logic}
  \label{algo:feedback_scout}
  \begingroup
  \footnotesize  
  \begin{flushleft}
    \textbf{Global:} weight vector $w$, cache of good paths $\mathit{buffer_{paths}}$, ECN count vector $ecn\_counts$\\
    \textbf{Input:} $feedback$ for specific path ($EV$, $i$ of $EV$ in $paths$), $\mathit{ecn_{rate}}$\\
  \end{flushleft}
  
  \begin{algorithmic}[1]
    \State \textbf{
      --------------------------------- Feedback Logic  ---------------------------------}
    \If{$feedback$ is ACK (no ECN)}
      \If{size of $\mathit{buffer_{paths}} < 8$ and $EV \notin \mathit{buffer_{paths}}$}
        \State Insert $EV$ into $\mathit{buffer_{paths}}$ based on latency
      \EndIf
    \EndIf
    
    \If{$feedback$ is ACK (ECN)}
      \State $ecn\_counts[i] \gets ecn\_counts[i] + 1$
      \If{$ecn\_counts[i] > ecn\_threshold$}
        \State $ecn\_counts[i] \gets 0$
        \State Remove $EV$ from $\mathit{buffer_{paths}}$
      \EndIf
    \EndIf
    
    \If{$feedback$ is NACK}
      \State $ecn\_counts[i] \gets 0$
      \State Remove $EV$ from $\mathit{buffer_{paths}}$
    \EndIf
    
    \If{$feedback$ is Timeout}
      \State $ecn\_counts[i] \gets 0$
      \State Remove $EV$ from $\mathit{buffer_{paths}}$
      \State $w[i] \gets 0$ \Comment Temporarily block path
    \EndIf
    \If{$\mathit{ecn_{rate}} > 90\%$}
      \State $w[0] \gets min\_bias\_factor$ \Comment Bias towards minimal paths
    \EndIf
  \end{algorithmic}
  \endgroup
\end{algorithm}

\paragraph{Tradeoff} \lbnamed-Scout balances exploration and stability. In the initial phase, it explores multiple paths using weighted sampling, which may introduce packet reordering but allows fast discovery of good paths. Once such paths are identified, the algorithm reuses them for as long as possible (no negative feedback). This reduces reordering but is less reactive to changing conditions. While newer transport protocols (UE~\cite{ultraethernet2025spec}, SRD~\cite{srd}, Falcon~\cite{10.1145/3718958.3754353}) are increasingly allowing out-of-order packet delivery to boost multipath performance, several legacy transports still require minimal to no re-ordering to work properly, making Spritz-Scout a valid middle ground.

\subsection{\lbnamed-Spray}\label{sec:algo_spray}


\begin{algorithm}[t]
  \caption{\lbnamed-Spray - Feedback Logic}
  \label{algo:feedback_spray}
  \begingroup
  \footnotesize 
  \begin{flushleft}
    \textbf{Global:} weight vector $w$, cache of good paths $\mathit{buffer_{paths}}$\\
    \textbf{Input:} $feedback$ for specific path ($EV$, $i$ of $EV$ in $paths$), $\mathit{ecn_{rate}}$\\
  \end{flushleft}
  
  \begin{algorithmic}[1]
    \State \textbf{
      --------------------------------- Feedback Logic  ---------------------------------}
    \If{$feedback$ is ACK (no ECN)}
      \If{size of $\mathit{buffer_{paths}} < 8$}
        \State Push $EV$ at the end of $\mathit{buffer_{paths}}$
      \EndIf
    \EndIf
    
    \If{$feedback$ is Timeout}
      \State $w[i] \gets 0$ \Comment Temporarily block path
    \EndIf

    \If{$\mathit{ecn_{rate}} > 90\%$}
      \State $w[0] \gets min\_bias\_factor$ \Comment Bias towards minimal paths
    \EndIf
  \end{algorithmic}
  \endgroup
\end{algorithm}

Algorithm~\ref{algo:feedback_spray} shows the feedback logic used in \lbnamed-Spray. The logic is simpler and has lower overhead compared to \lbnamed-Scout. Good paths are pushed to the end of the circular buffer $\mathit{buffer_{paths}}$ upon positive feedback. Paths will be added even if already present, since each use removes the corresponding EV from the buffer in the send logic (Algorithm~\ref{algo:send_both}). ECN marks and NACKs are ignored. On timeout and heavy ECN congestion we rely on the same mechanism as in \lbnamed-Scout.

\subsection{Interoperability and Fault Tolerance}\label{sec:loadbalancing_inter}

\paragraph{Congestion Control}
\sloppy{
\lbname is compatible with sender-based congestion control schemes. Since low-diameter topologies include paths with varying round trip times, we recommend using ECN-based congestion control that supports per-packet acknowledgments. One such example is DCTCP~\cite{alizadeh2010data} as implemented in MPRDMA~\cite{lu2018multi}. For the evaluation, we use an extended version of DCTCP that incorporates the QuickAdapt and FastIncrease techniques from SMaRTT~\cite{bonato2026smarttsenderbasedmarkedrapidlyadapting}, enabling fast and precise congestion window adjustments.
}

\paragraph{Failures}
\lbname relies on source-guided adaptive routing, which enables fast reaction to both transient and persistent failures. EV entries can be marked as temporarily or permanently unavailable, allowing the sender to avoid paths affected by link or switch failures without relying on in-network recovery. 

\paragraph{Packet Trimming} 
In this work, we assume the use of packet trimming to detect losses. This approach not only simulates the latest capabilities of network switches (UE recommends the use of trimming) but also simplifies our analysis. Packet trimming functions as an early loss notification mechanism; without it, feedback on packet loss would be delayed. For example, if a packet is dropped at a switch due to a full queue, packet trimming immediately informs the sender of this event, whereas alternative methods like timeout-based detection would result in a significant notification delay. While such delays could adversely affect overall performance, this impact would be consistent across all load balancing and congestion control algorithms, as demonstrated in \cite{bonato2026smarttsenderbasedmarkedrapidlyadapting}.

\paragraph{ACK Coalescing}
Throughout this work, and consistently with the latest capabilities of modern interconnects such as Slingshot \cite{de2020depth}, we assume the use of per-packet acknowledgments. This design ensures that the sender is continuously supplied with up-to-date information on the network status, thanks to ECN or delay-based feedback. Although this approach introduces some additional overhead, ACK packets are typically very small compared to modern MTUs (4 or 8 KiB), resulting in an overall overhead of approximately 1\%.

With limited ACK rates, Spritz-Scout is largely unaffected because it retains buffered elements. Spritz-Spray degrades as ACK coalescing grows \cite{bonato2025repsrecycledentropypacket}, yet it still delivers gains in steady state and preserves them under failures.
\section{Evaluation}\label{sec:evaluation}

In this section, we evaluate the performance of \lbnamed-Scout and \lbnamed-Spray against established routing and sender-based load balancing schemes. Experiments are conducted on both Dragonfly and Slim Fly. We extend \textit{htsim}~\cite{handley2017re}, a packet-level simulator, to support low-diameter topologies, Source-Guided Adaptive Routing, and the Spritz framework.

\subsection{Simulation Setup}\label{sec:evaluation_setup}

\paragraph{Network Setup}
We construct networks with approximately 1000 endpoints for both Dragonfly and Slim Fly. We assume modern high-performance Ethernet components. All network construction parameters and hardware details are summarized in Table~\ref{tab:topo_params}. The bandwidth delay product (BDP) is calculated based on the longest path for each topology.

We model a lossy Ethernet environment consistent with the Ultra Ethernet Transport specification: queues may ECN-mark, trim payloads, or drop packets. Senders utilize a DCTCP-style ECN control loop, reacting to trimming feedback when available, and reverting to timeouts for packet loss due to failures or corruption. Note that by design, Spritz is largely orthogonal to the specific loss recovery mechanism.

\paragraph{Congestion Control}
We use an extended version of DCTCP that includes QuickAdapt and FastIncrease optimizations, as described in Section~\ref{sec:loadbalancing_inter}. This setup reflects the latest state-of-the-art in sender-based congestion control, using only the ECN signal to react to congestion, which is preferred due to variable-length paths in low-diameter topologies. The same congestion control is enabled for all evaluated schemes. The initial and maximum congestion window is set to 1.5 times the BDP, calculated based on the longest path in each topology.

\begin{table}[t]
\centering
\caption{Simulation parameters}
\begingroup
\setlength{\tabcolsep}{4pt}      
\renewcommand{\arraystretch}{0.9}
\scriptsize                     
\begin{tabular}{l|lllll}
\toprule
\multicolumn{6}{c}{\textbf{Topology Parameters}} \\
\midrule
Dragonfly & $p=4$ & $a=8$ & $h=4$ & 1056 endpoints & 264 switches \\
Slim Fly & $p=7$ & $q=9$ & & 1134 endpoints & 162 switches \\
\midrule
\multicolumn{6}{c}{\textbf{Network Parameters}} \\
\midrule
Link Speed & \multicolumn{5}{l}{400 Gbps} \\
Local Links & 25 ns & & \multicolumn{3}{l}{(5 ns/m\textsuperscript{\ref{footnote:cable}}, assume 5 m)} \\
Global Links & 500 ns & & \multicolumn{3}{l}{(5 ns/m\textsuperscript{\ref{footnote:cable}}, assume 100 m)} \\
Switch Latency & \multicolumn{5}{l}{500 ns} \\
Queue Size & \multicolumn{2}{l}{$1 \cdot BDP$} & \multicolumn{3}{l}{(per port, DF: 88 pkts, SF: 92 pkts)} \\
ECN $K_{min}$, $K_{max}$ & \multicolumn{5}{l}{0.2, 0.8} \\
Packet Size & \multicolumn{5}{l}{64 B (header) + 4096 B (payload)} \\
\midrule
\multicolumn{6}{c}{\textbf{\lbnamed}} \\
\midrule

Explore Threshold & 44 packets & & \multicolumn{3}{l}{($\sim 0.5 \cdot BDP$)} \\
ECN Threshold & 8 packets & & \multicolumn{3}{l}{($\sim 0.1 \cdot BDP$)} \\
\bottomrule
\end{tabular}
\endgroup
\label{tab:topo_params}
\vspace{-1em}
\end{table}
\footnotetext[1]{\label{footnote:cable}Based on \href{https://docs.nvidia.com/cabling-data-centers.pdf}{NVIDIA Cabling Data Centers Design Guide}.}

\paragraph{Baselines}
We group all compared schemes into four categories, based on routing and sender-based load balancing granularity.

\textbf{Routing.} We use Minimal, Valiant, and UGAL-L, all with switch-level decisions. In UGAL-L, the switch selects the minimal path if \(q_{\min}h_{\min} \le q_{\mathrm{val}}h_{\mathrm{val}}\), where \(q_{\min}\) and \(h_{\min}\) are the queue occupancy and hop count of the minimal path, and \(q_{\mathrm{val}}\) and \(h_{\mathrm{val}}\) are those of a randomly chosen valiant path.

\textbf{ECMP} Each flow is assigned to a fixed path based on its five-tuple (Section~\ref{sec:background_topologies}). Assigned paths may be minimal or non-minimal. The assignment is static and does not adapt to congestion.

\textbf{Coarse-Grain Sender-Based Load Balancing} This includes prior sender-side load balancing schemes at flowlet or flowcell granularity. These schemes aim to reduce packet reordering while still reacting to congestion feedback. We implement \textit{Flicr}, the most relevant scheme in this category, as it targets directly connected networks~\cite{kabbani2017flier}.

\textbf{Fine-Grain Sender-Based Load Balancing} This includes packet spraying techniques enabled by the Spritz framework. We implement and evaluate three schemes: Oblivious Packet Spraying (OPS)~\cite{spraying}, \lbnamed-Scout (Section~\ref{sec:algo_scout}), and \lbnamed-Spray (Section~\ref{sec:algo_spray}). 

We consider both uniform (u) and weighted (w) versions of most packet spraying schemes. In the weighted version, all shorter paths are assigned weights based on their latency, as described in Section~\ref{sec:routing} (Equation~\ref{eqn:weight}). These weights are further scaled for evaluation by a factor of 3 to increase the selection probability of shorter paths. Longest paths have always weight 1.0.

Finally, we note that the original definition of Valiant routing \cite{valiant1982scheme}
assumes uniform probability across all paths, making it conceptually closer to OPS (u) than to our implementation of the Valiant routing where the next hop is
sampled uniformly but independently at each switch. As a result, whenever a global hop is necessary, paths using the direct global links will have higher probability compared to the other global links from the same group. For completeness we keep both approaches in the results.


\subsection{Workloads}\label{sec:evaluation_workloads}

\begin{figure}[!t]
\centering
\includegraphics[width=1.0\linewidth]{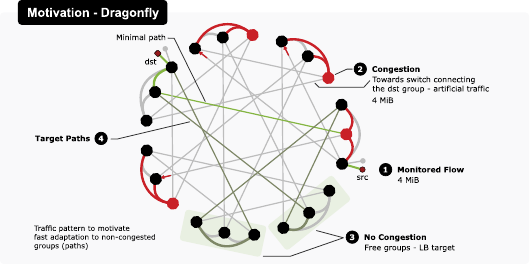}
\caption{
Illustration of a motivational example showing fast adaptation to non-congested groups (paths) in a Dragonfly topology.
\includegraphics[scale=0.6,trim=0 1 0 0]{figures/icons/Circle1Icon.pdf}~We observe the flow completion time (FCT) of a single monitored 4 MiB flow (\includegraphics[scale=0.6,trim=0 -2 0 0]{figures/icons/MinPathIcon.pdf}) between a specific source \includegraphics[scale=0.6,trim=0 1 0 0]{figures/icons/SrcIcon.pdf} and destination \includegraphics[scale=0.6,trim=0 1 0 0]{figures/icons/DstIcon.pdf} endpoints.
\includegraphics[scale=0.6,trim=0 1 0 0]{figures/icons/Circle2Icon.pdf}~Most groups are heavily congested: many flows target a switch with a global link to the destination group, creating significant queue buildup.
\includegraphics[scale=0.6,trim=0 1 0 0]{figures/icons/Circle3Icon.pdf}~Few groups do not have this traffic and thus congestion (free groups).
\includegraphics[scale=0.6,trim=0 1 0 0]{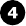}~The goal of the load balancing scheme is to quickly discover and use these target paths.
}
\label{fig:evaluation:motivation}
\vspace{-1em}
\end{figure}

\begin{figure*}[!ht]
\centering
\includegraphics[width=1.0\textwidth]{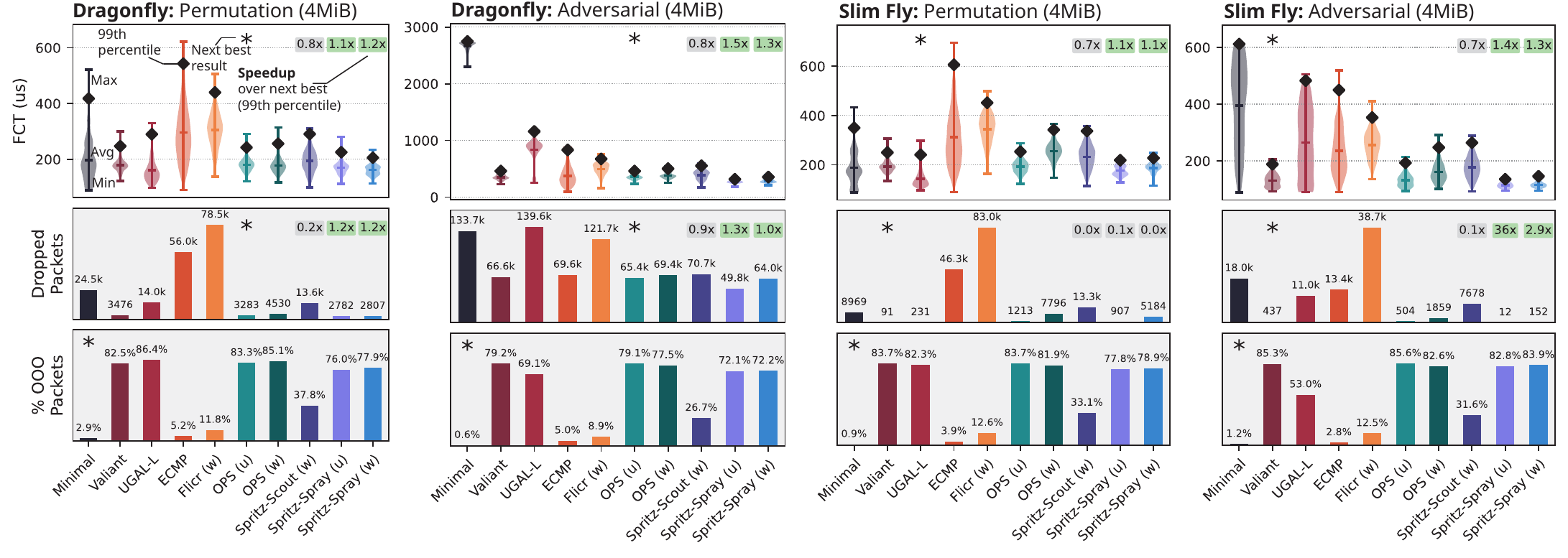}
\caption{Distribution of flow completion times (FCTs), number of dropped packets, and percentage of packets arriving out-of-order (OOO) on microbenchmarks (permutation and adversarial patterns) for Dragonfly and Slim Fly. OOO packets in ECMP and Minimal result from retransmissions due to congestion drops.}
\label{fig:eval_microbench}
\vspace{-1em}
\end{figure*}

\begin{figure*}[!ht]
\centering
\includegraphics[width=1.0\textwidth]{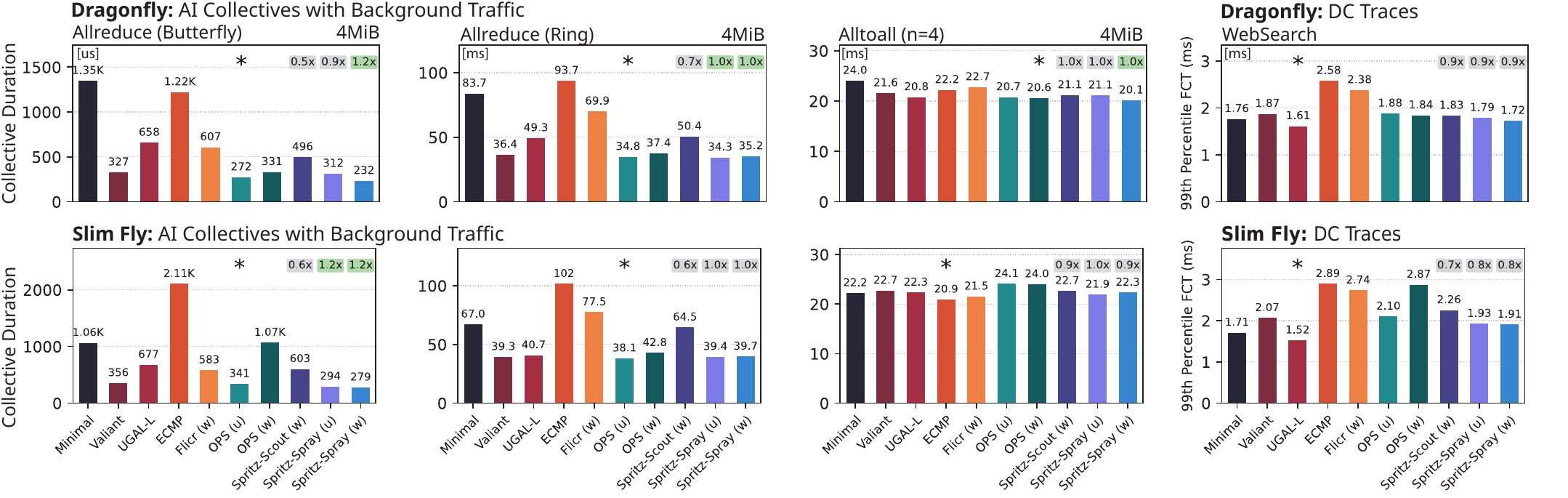}
\caption{
Duration of AI collectives in shared environment (background permutation ECMP traffic). 99th percentile of flow completion times (FCTs) of web search datacenter workload (traces). All experiments are evaluated on both Dragonfly and Slim Fly.
}
\label{fig:eval_collectives}
\vspace{-1em}
\end{figure*}

\paragraph{Microbenchmarks}
We start by showcasing an artificial adversarial pattern that showcases the strengths of Spritz. In this case we monitor the flow completion time (FCT) of a 4~MiB flow between a chosen source and destination pair while inducing congestion in nearly all groups, including the source, by launching background flows to switches that connect each congested group to the destination group. Because routing relies only on local information, the sender and local switches cannot identify which groups are congested, so the monitored flow often traverses congested groups and incurs additional queuing delay and possible packet drops. A few groups, including the destination group, remain idle (Figure~\ref{fig:evaluation:motivation} gives more details about the scenario).

We then shift to two synthetic traffic patterns commonly used to study routing performance in low-diameter topologies. The first is the permutation pattern, where each endpoint sends to one randomly chosen receiver, ensuring that every endpoint communicates with exactly one other. We further prioritize the receiver to be outside the local group to force utilization of global links. The second is the adversarial pattern, which is specific to each topology~\cite{kim2008technology, besta2014slim}. It is designed to stress the network by overloading specific links, exposing cases where minimal routing performs poorly. We exclude pure incast as a primary benchmark because it is dominated by the receiver bottleneck, effectively testing congestion control rather than routing efficiency. However, because incast creates widespread congestion that degrades unrelated traffic, we instead evaluate a mixed workload (incast plus bystanders) to quantify this collateral damage on the bystanders.

\paragraph{AI Collectives}
We use two communication collectives widely used in AI training: Allreduce~\cite{desensi2024swingshortcuttingringshigher, li2020pytorch} and Alltoall~\cite{naumov2019deep}. Allreduce is a key primitive in data parallelism and tensor parallelism for large language model training. Alltoall is relevant for mixture-of-experts models. We implement Allreduce using both the ring and butterfly algorithms. For Alltoall we limit each endpoint to $n$ parallel connections.

To better reflect a shared datacenter environment and to keep simulation time manageable, we run the collectives on a subset of 128 endpoints. All other endpoints in the network participate in a permutation among themselves using ECMP as load balancing. This creates additional load and mimics the presence of other jobs in the system or an incremental deployment scenario for Spritz.

\paragraph{Datacenter Traces}
We use a web search workload trace collected from a production datacenter, previously used in related work~\cite{alizadeh2010data, vanini2017let}. To avoid incast, we randomly select receivers while limiting the number of simultaneous senders per receiver. This further spreads the connections uniformly across the network. We run the workload at full load (1.0) for a duration of 1 ms.

\subsection{Results}\label{sec:evaluation_results}
\begin{table}[t]
\centering
\caption{FCT of the monitored flow for each scheme. The last three columns indicate the Spritz variants.}
\label{tab:evaluation:motivation}
\setlength{\tabcolsep}{2.5pt}       
\renewcommand{\arraystretch}{0.95}  
\scriptsize
\resizebox{\columnwidth}{!}{%
\begin{tabular}{@{}l*{10}{c}@{}}
\toprule
& \rotatebox[origin=c]{90}{Minimal} 
& \rotatebox[origin=c]{90}{Valiant} 
& \rotatebox[origin=c]{90}{UGAL-L} 
& \rotatebox[origin=c]{90}{ECMP} 
& \rotatebox[origin=c]{90}{Flicr (w)} 
& \rotatebox[origin=c]{90}{OPS (u)} 
& \rotatebox[origin=c]{90}{OPS (w)} 
& \textbf{\rotatebox[origin=c]{90}{Scout}} 
& \textbf{\rotatebox[origin=c]{90}{Spray (u)}} 
& \textbf{\rotatebox[origin=c]{90}{Spray (w)}} \\
\midrule
FCT [$\mu$s] (solo) & 91 &  &  &  &  &  &  &  &  &  \\
FCT [$\mu$s]        & 244 & 228 & 199 & 502 & 227 & 187 & 173 & \textbf{110} & \textbf{121} & \textbf{113} \\
Speedup vs UGAL-L   &     &     & $*$ &     &     &     &     & \textbf{1.8$\times$} & \textbf{1.6$\times$} & \textbf{1.8$\times$} \\
\bottomrule
\end{tabular}%
}
\vspace{-1em}
\end{table}
We now present the results of our evaluation. Figure~\ref{fig:eval_microbench} shows the distribution of FCTs, the number of dropped packets, and the percentage of packets arriving out-of-order (packet is OOO if $ \text{PSN}_{\text{exp}} \neq \text{PSN}_{\text{recv}} $) for the microbenchmarks. Figure~\ref{fig:eval_collectives} shows the duration of AI collectives (Allreduce butterfly, Allreduce ring, Alltoall) executed by a randomly selected group of 128 endpoints in a shared environment. It further shows the 99th percentile FCTs for the web search datacenter workload. All experiments are evaluated on both Dragonfly and Slim Fly topologies.

\paragraph{Microbenchmarks}
In the microbenchmark proposed in Figure~\ref{fig:evaluation:motivation} UGAL-L adapts to queue occupancy but lacks visibility into which output ports reach free groups, so it can choose global links that look lightly loaded yet lead into congested groups; with few free groups, it rarely finds the desired paths. In contrast, Spritz-Scout and Spritz-Spray quickly discover and exploit target paths, achieving up to 1.8x speedup over UGAL-L as shown in Table~\ref{tab:evaluation:motivation}.

In the other classical microbenchmarks Spritz-Spray consistently outperforms all other schemes, including adaptive routing (UGAL-L), across all experiments. The speedup of Spritz over next best is around 1.1-1.2x for permutations, while for the adversarial patterns it is around 1.3-1.5x. It also results in the fewest packet drops in all cases except the adversarial pattern on Slim Fly, where it ranks third best but remains competitive. As expected, minimal routing and UGAL-L perform poorly on the adversarial pattern in both topologies due to overloaded minimal paths.

\begin{figure}[!t]
\centering
\includegraphics[width=1.0\linewidth]{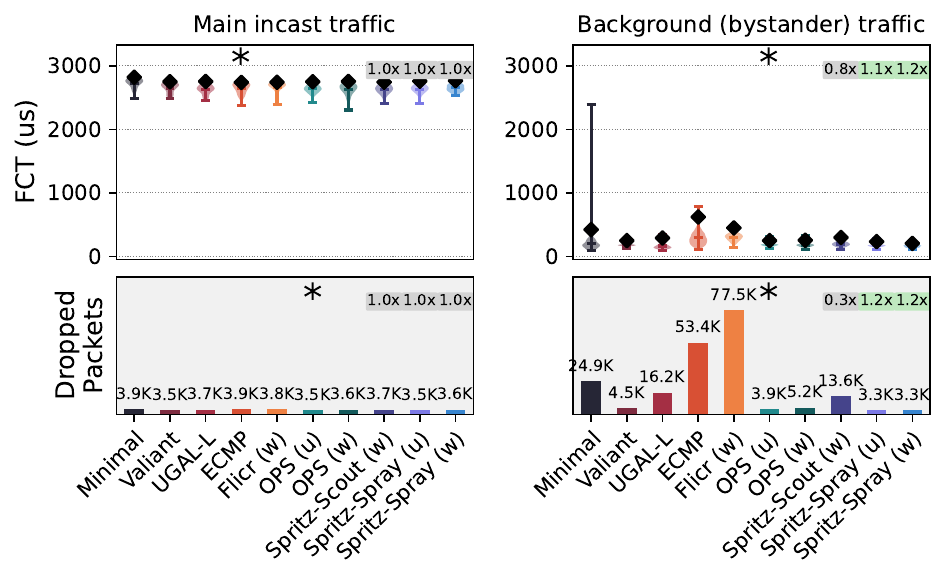}
\caption{
32x4 MiB incast to node 160 with a disjoint 1023-flow permutation background (all start at 0). Violin shows the bystander FCT distribution; bars show retransmissions (loss proxy). Spritz-family source selection improves bystander tail latency by avoiding persistent coupling into the incast congestion region.}
\label{fig:incast_perf}
\vspace{-1.5em}
\end{figure}

Incast + Bystanders microbenchmark. To test interference beyond Alltoall patterns, we evaluate a mixed workload combining a synchronized incast hotspot with concurrent background flows. On Dragonfly (1056 nodes), 32 senders (nodes 0-31) each send a 4 MiB flow to a single receiver (node 160), while 1023 additional 4 MiB bystander flows form a disjoint one-to-one permutation; all flows start at time 0. Results are shown in Figure \ref{fig:incast_perf} where, as expected, incast performance is not the differentiator: incast p99 FCT is similar across schemes ($\approx$2.74--2.77 ms; Minimal slightly worse at $\approx$2.82 ms). In contrast, bystander performance varies substantially depending on how much each scheme couples bystanders into the hotspot region. The best baseline (Valiant) achieves bystander p99 FCT of 249.01 $\mu$s, while the best Spritz-family variant reduces it to 204.40 $\mu$s ($-17.9\%$) and lowers bystander retransmissions (4526 $\rightarrow$ 3323). In summary, Spritz mitigates collateral damage to bystander traffic rather than resolving the incast bottleneck itself, which remains fundamentally limited by congestion control.

\paragraph{AI Collectives}
Spritz achieves fast completion times and low number of dropped packets (not shown here due to space constraints) across all evaluated collectives. It consistently performs best overall or the best among packet spraying techniques. One exception is Alltoall on Slim Fly, where it remains competitive with other packet spraying schemes. A similar trend is observed for the number of dropped packets.

\begin{figure*}[!t]
\centering
\includegraphics[width=1.0\textwidth]{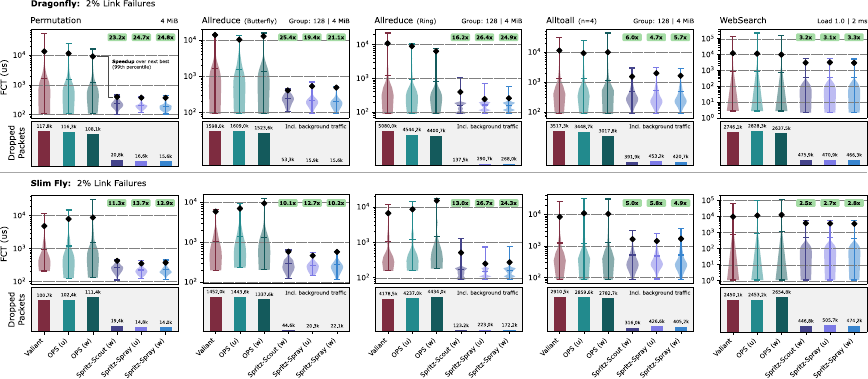}
\caption{
Distribution of flow completion times (FCTs) in microbenchmarks, AI collectives in shared environment (background traffic), and datacenter traces for Dragonfly and Slim Fly with 2\% of fully failed links. Minimal, UGAL-L, ECMP, and Flicr were not able to finish the traffic in the time limit (1 s). 
}
\label{fig:eval_fail}
\vspace{-1.1em}
\end{figure*}

\paragraph{Datacenter Traces}
Spritz achieves the lowest number of dropped packets and performs best among spraying schemes. However, packet spraying techniques show higher FCTs compared to minimal and UGAL-L, especially on Slim Fly. This is due to the uniform nature of the workload, which favors consistent minimal-path routing and the large number of tiny flows where UGAL-L shines, since source-based schemes are reactive. We note that if the traffic pattern is known to consistently contain mostly tiny flows, then it would be beneficial to increase the minimal paths bias. In this work we keep identical weights, even in Spritz-unfavorable cases.


\subsection{Failures}\label{sec:evaluation_failures}

So far, we evaluated performance under ideal network conditions. While Spritz showed strong improvements, other packet spraying schemes also performed well. We now study resilience to link failures across permutation traffic, AI collectives, and datacenter traces. To simulate failures, we randomly disable 2\% of the links in the network.

We compare only against baselines capable of adapting to network changes. We exclude Minimal, UGAL-L, ECMP, and Flicr. Minimal and ECMP require switch-level support to reroute around failures. In practice, failure detection and forwarding table updates can take several seconds, leading to significant packet loss~\cite{bonato2025repsrecycledentropypacket}. UGAL-L is limited to local decisions and may not detect failures beyond its immediate view, for example in remote groups. Flicr reacts slowly and, in our experiments, failed to deliver all packets within the time limit (1s). Therefore, the main baseline for this case is Valiant and both versions of OPS (uniform and weighted).

In Figure \ref{fig:eval_fail} Spritz demonstrates its strong ability to react quickly to link failures. This fast adaptation leads to consistent speedups ranging from 2.5x to 25.4x over the next best. This is further demonstrated by the number of dropped packets, which is in some cases two orders of magnitude lower compared to Valiant and OPS. The results highlight the strength of Spritz in handling dynamic network conditions with minimal losses.

\section{Related Work}\label{sec:relatedwork}

An established class of low-diameter topologies reduces cost and power consumption compared to traditional networks such as Fat Trees or Torus. Numerous topologies belong to this class, including {Slim Fly}~\cite{besta2014slim, blach2024high}, Dragonfly~\cite{kim2008technology}, {PolarFly}~\cite{lakhotia2022polarfly, lakhotia2023network}, PolarStar~\cite{lakhotia2024polarstar}, {Xpander}~\cite{valadarsky2015}, {BundleFly}~\cite{lei2020bundlefly}, Spectralfly~\cite{young2022spectralfly}, and numerous other designs also in the on-chip~\cite{besta2018slim, gianinazzi2022spatial} and chiplet settings~\cite{iff2023hexamesh, 10.1109/DAC56929.2023.10247754, iff2023rapidchiplet, iff2025foldedhexatorus}. These topologies are important as they have achieved new frontiers in performance, cost, and power consumption. While we focus on Slim Fly and Dragonfly in this paper, these networks were shown to have similar properties in terms of path diversity~\cite{besta2020fatpaths}. Hence, Spritz could be easily used with other low-diameter networks.

Traditionally, low-diameter networks use routing strategies at the packet-level with the routing decision done at the switch level. This involves: minimal, Valiant, and UGAL-L routing~\cite{kim2008technology, besta2014slim}. Several other variations or improvements of such schemes have been proposed~\cite{besta2020high, benito2018analysis, chaulagain2024enhanced, benito2019acor}. Schemes, such as Flicr (also known as Flier)~\cite{kabbani2017flier} focus on general directly connected networks at flowlet or flowcell level, but do not discuss integration with low-diameter topologies. 

In contrast, routing and load balancing strategies for datacenter networks based on fat tree topologies feature a wider range of designs and techniques. Packet-level spraying schemes that aim to maximize performance by distributing packets across multiple paths include: OPS (also known as RPS)~\cite{spraying}, REPS~\cite{bonato2025repsrecycledentropypacket}. Flow, flowlet, and flowcell level schemes that focus on minimizing packet reordering while offering varying levels of adaptivity include: standard ECMP~\cite{hopps2000analysis}, WCMP~\cite{zhou2014wcmp}, Flowlet Switching~\cite{vanini2017let, kandula2007dynamic}, Presto~\cite{he2015presto}, CONGA~\cite{alizadeh2014conga}, FlowBender~\cite{kabbani2014flowbender}, Flicr~\cite{kabbani2017flier}, and PLB~\cite{qureshi2022plb}.

Spritz targets high-performance Ethernet infrastructures. However, thanks to its simplicity, it could easily be implemented in other architectures, using, for example, programmable SmartNICs and ideas such as sPIN~\cite{hoefler2017spin, di2019network, di2022building}.

However, many of these schemes do not account for variable-length paths or consider low-diameter topologies. Switch-based routing schemes such as UGAL-L rely on local information, which can limit their effectiveness. Schemes that incorporate global network conditions often rely on vendor-specific hardware, making them difficult to deploy in environments that prioritize commodity, vendor-neutral switches.
Prior work explored adapting UGAL/UGAL-L’s minimal vs. non-minimal bias using runtime feedback and local counters to better reflect global congestion conditions. More recently, learning-based approaches have been proposed to approximate UGAL-G like decisions using only locally observable signals (e.g., UGAL-ML \cite{ugalml}), and reinforcement-learning policies have been studied for Dragonfly adaptive routing \cite{ml3}. While these approaches can be effective, Spritz offers a distinct advantage: a simpler, endpoint-driven control loop that adapts rapidly to network changes, particularly under failure conditions.
\section{Conclusion}\label{sec:conclusion}
Spritz is a sender-based load balancing framework designed for low-diameter topologies such as Dragonfly and Slim Fly. We shift routing decisions to the endpoints by introducing Source-Guided Adaptive Routing. This enables fine-grained control over routing with fast feedback, improving both performance and resilience. Spritz uses standard Ethernet features and does not rely on proprietary hardware support.

We introduce two specific algorithms, Spritz-Scout and Spritz-Spray. Through simulations on microbenchmarks, AI collectives, and realistic workloads, Spritz consistently delivers low flow completion times and fewer packet drops. It also shows a strong ability to respond quickly to link failures, maintaining stable performance when the network degrades.

Spritz is practical to deploy, compatible with existing Ethernet infrastructure, and well-suited for modern datacenters. It enables efficient use of low-diameter topologies offering a flexible and scalable path forward for building high-performance, vendor-neutral networks in the Ultra Ethernet era.

\section*{Acknowledgments}
 This project received funding from the European Research Council (ERC) under the European Union’s Horizon 2020 research and innovation program (grant agreement PSAP, No. 101002047) and a CAF America grant. This work is partially supported by Broadcom's VMware University Research Fund. We also thank the Swiss National Supercomputing Center (CSCS) for providing the computational resources used in this work. ChatGPT assisted with editing and quality control.

\bibliographystyle{IEEEtran}
\bibliography{references}

\end{document}